\documentclass[pra,aps,a4paper,twocolumn,repreprint,grouppedaddress,longbibliography]{revtex4}

\usepackage{amssymb}
\usepackage{amsmath}
\usepackage{amsfonts}
\usepackage{graphicx}
\usepackage{color, soul}
\usepackage{natbib}
\usepackage{hyperref}

\def\be{\begin{equation}}
\def\ee{\end{equation}}
\newcommand{\eps}{\varepsilon}
\def\bq{\mathbf{q}}

\begin{document}
	
	\title{Local thermal fluctuations in current-carrying superconducting nanowires}
	
	\author{Alexej D. Semenov}
	\affiliation{German Aerospace Center (DLR), Institute of Optical Sensor Systems, Rutherfordstra$\beta$e 2, 12489 Berlin, Germany}
	
	\author{Mariia Sidorova}
	\affiliation{German Aerospace Center (DLR), Institute of Optical Sensor Systems, Rutherfordstra$\beta$e 2, 12489 Berlin, Germany}
	
	\author{Mikhail A. Skvortsov}
	\affiliation{Skolkovo Institute of Science and Technology, 121205 Moscow, Russia}
	\affiliation{L. D. Landau Institute for Theoretical Physics, 142432 Chernogolovka,
		Russia}
	
	\author{Artem Kuzmin}
	\affiliation{Karlsruher Institute of Technology (KIT), Institute of Micro- und Nanoelektronc Systems (IMS), 76187 Karlsruhe, Germany}
	
	\author{Konstantin Ilin}
	\affiliation{Karlsruher Institute of Technology (KIT), Institute of Micro- und Nanoelektronc Systems (IMS), 76187 Karlsruhe, Germany}
	
	\author{Michael Siegel}
	\affiliation{Karlsruher Institute of Technology (KIT), Institute of Micro- und Nanoelektronc Systems (IMS), 76187 Karlsruhe, Germany}

	\begin{abstract}
		We analyze the effect of different types  of fluctuations in internal electron energy on the rates of dark and photon counts in straight current-carrying superconducting nanowires. Dark counts appear due to thermal fluctuations in statistically independent cells with the effective size of the order of the coherence length; each count corresponds to an escape from the equilibrium state through an appropriate saddle point. For photon counts, spectral broadening of the deterministic cut off in the spectra of the detection efficiency can be phenomenologically explained by local thermal fluctuations in the electron energy within cells with the same effective volume as for dark counts.
	\end{abstract}
	
	\date{\today}
	\maketitle

	\section{Introduction}
	
	All kinds of second-order phase transitions are smeared by fluctuations. For example, superconducting fluctuations in the normal state according to Aslamazov and Larkin \cite{1} broaden the resistive transition of thin films. Microelectronic devices utilizing thin superconducting films also suffer fluctuations which deteriorate the performance metrics of these devices. Specifically, for superconducting nanowire single-photon detectors, it is believed that fluctuations are responsible for smearing otherwise deterministically sharp energy threshold (spectral cut-off) between photons which are surely detected and photons which are not detected in any way \cite{2,3}. Such a detector is a narrow albeit two-dimensional thin superconducting strip carrying the current less but close to the critical current. Photon absorbed in the strip gives rise to a nonequilibrium ``hot"  spot that reduces the current-carrying ability of the strip around the absorption site. If the photon energy is sufficient, the superconducting state breaks down locally. Energy dissipated in the resistive spot initiates growth of a normal domain which may have a length of a few times larger than the strip width. Ones the domain starts to grow, current diverts from the strip to the read out line. This causes the domain to shrink. When it disappears the current returns into the strip and after a dead time the strip is ready to detect another photon. This event produces a voltage transient which is called photon count. However the strip still generates counts even when it is not illuminated by light. Such counts are called dark counts.
	
	Perhaps the most important performance metrics of these detectors is the timing jitter that measures stochastic variations in the time delay between arrival of a photon and the appearance of the detector response to this photon \cite{4,5,6,7}. So far several effects which may cause jitter have been considered. Fano fluctuations \cite{8}randomize the portion of the energy of the absorbed photon which is delivered to electrons via the cascade of scattering events between electrons and phonons \cite{9}. The amount of the delivered energy directly affects the delay time that causes jitter. Another effect is the dependence of the delay time on the position of the photon absorption cite across the strip \cite{10}. The effect of fluctuations (local variations) in the fim thickness was theoretically studied in Ref.\ \cite{Cheng 2017}. All those fluctuations will also smear the deterministic spectral cut-off.
	
	Here we analyse the effect of different fluctuations on the experimental spectra of the photon count rate in straight superconducting strips. The straight strips instead of meanders were chosen in order to avoid bends, which are known to dominate in the dark count rate and in the photon count rate at small photon energies \cite{11}. We first study statistics and rate of dark counts as a function of current and temperature in straight superconducting strips and evaluate the involved energy barrier and the length scale.  We fürther show that fluctuations discussed in literature do not quantitatively explain experimental spectral broadening in the photon count rate. We then invoke local thermal fluctuations and estimate the optimal volume for these fluctuations which is required to decribe quantitatively experimental spectral broadening.
	
	The paper is organized as follows. Section \ref{S2} is devoted to the specimens and experimental results. In Sec.\ \ref{S3} we examine fluctuation sources for their capability to fit experimental data. Section \ref{S4} contains the discussion and concludes the paper. Mathematical details and the results of the AFM study of film thickness are collected in the Appendixes \ref{A} and \ref{B}, respectively.

	%%%%%%%%%%%%%%%%%%%% Figure/Image No: 1 starts here %%%%%%%%%%%%%%%%%%%%
	
	\begin{figure}
		\centerline{\includegraphics[width=0.48\textwidth]{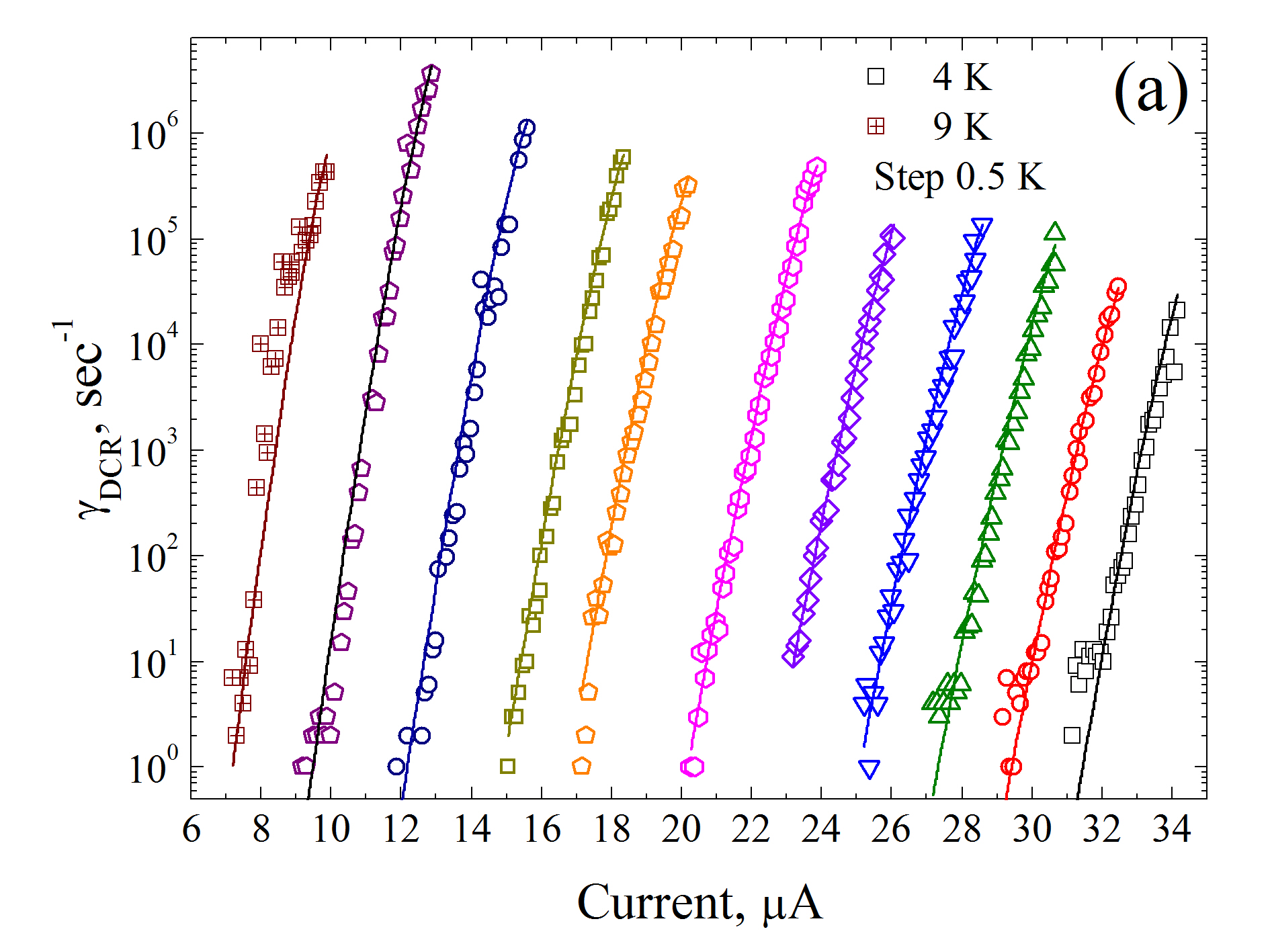}}
		\centerline{\includegraphics[width=0.48\textwidth]{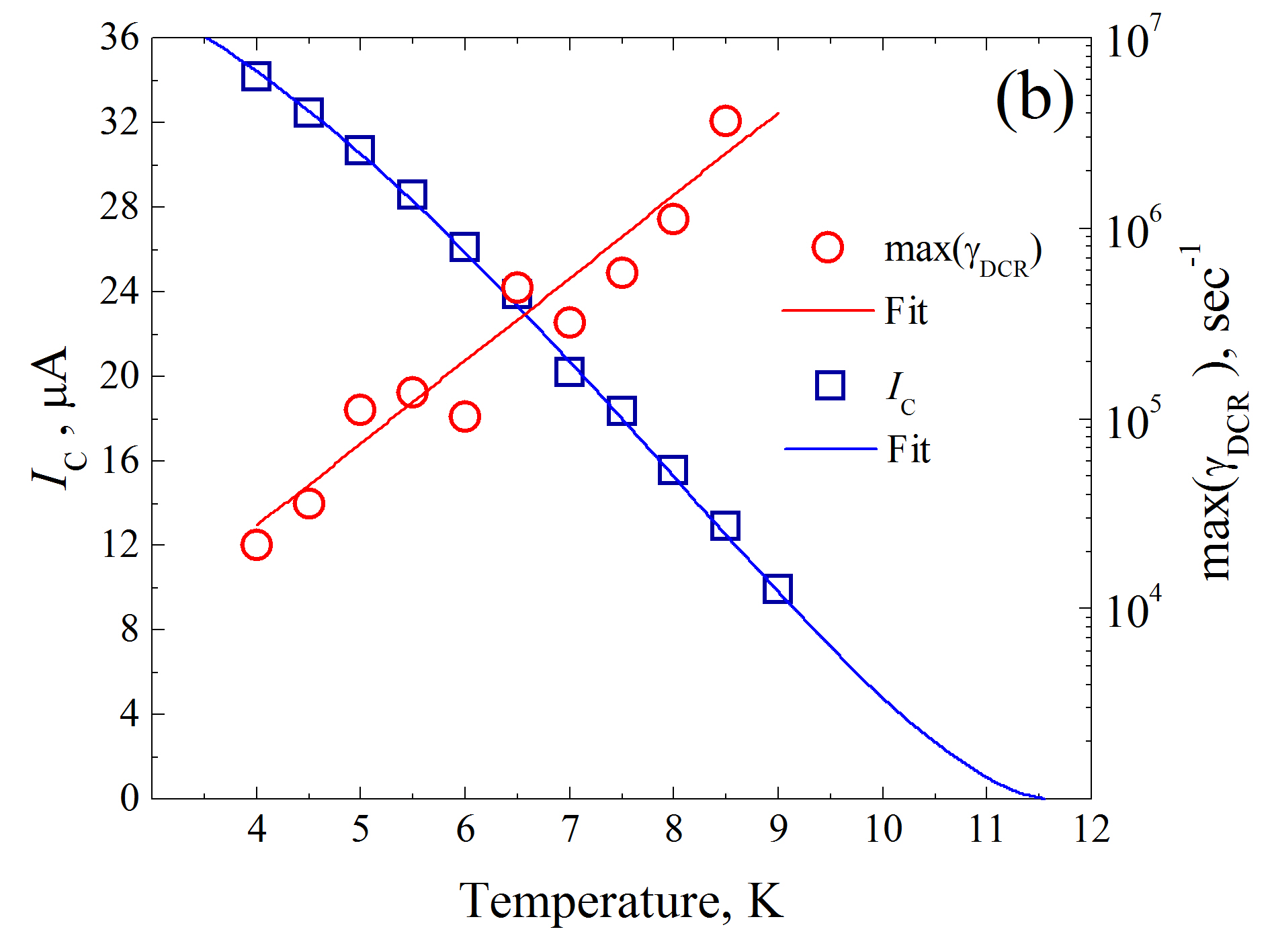}}
		\caption{(a) Dark count rates as functions of current at several temperatures specified in the legend. Solid lines represent fits with the generalized 2D LAMH model (\ref{eq1}). (b) Maximum rates of dark counts and critical current as functions of temperature. Solid lines represent fits described in the text.}
		\label{F1}
	\end{figure}
	
	%%%%%%%%%%%%%%%%%%%% Figure/Image No: 1 Ends here %%%%%%%%%%%%%%%%%%%%	

\section{Experiment}
	\label{S2}
	
	\subsection{Specimens and experimental details}
	
	We used straight narrow superconducting NbN strips which were geometrically identic and had a length of 40 $\mu$m and a width of 100 nm. NbN films with a thickness of 5 nm were deposited on Al$_2$O$_3$ substrates by reactive magnetron sputtering. The strips were drawn by the electron-beam lithography with negative Polymethylmethacrylate resist, which results in significant improvement of the superconducting characteristics \cite{12}. The active strip was surrounded by parallel equally spaced and electrically suspended strips of the same width in order to eliminate diffraction and to obtain uniform optical coupling. The active strip was imbedded via tapers in the middle of the much wider central line of a few millimeter long coplanar waveguide (CPW). The CPW was electrically short cut at the one side of the strip. Voltage transients were acquired at the end of the CPW on the other side of the strip. More details on the layout of specimens are presented in Ref. \cite{5}.
	
	A few studied specimens had almost identical parameters. From the fit of the resistive transition with the Aslamazov-Larkin fluctuation model \cite{1}, we found the mean-field transition temperature 11.8 K. Transport measurements showed a critical current of 34 $\mu$A at 4.2 K and the normal sheet resistance 330 $\Omega/\Box$ at 25 K.
	
	Specimens were mounted either in the deep stick with optical access trough the single mode fiber or in the optically tight continuous flow cryostat. The former setup was used to acquire spectra of the single-photon response at 4.5 K and different bias currents, while the latter served for study of dark counts in the temperature range from 4.5 to 9 K. Fiber in the former setup was \textit{in situ}\/ aligned against the active strip to maximize the detection efficiency. This ensures uniform illumination across the strip. For spectral measurements we used a monochromator delivering light in the wavelength range from 350 to 2500 nm with the spectral resolution from 0.012 to 0.1 eV, respectively, and three continuous lasers at the wavelengths 532, 633 and 1064 nm. The coplanar line with the specimen was connected to the coaxial cable that guided voltage transients to the warm amplifier outside of the deep stick or cryostat. The coaxial bias tee was plugged in the coaxial line right before the amplifier. Counts delivered by the specimen were registered either by the real-time oscilloscope or by the computer controlled TCSPC (time correlated single photon counting) card.

	\subsection{Mean dark count rate}
	
	Current dependences of the mean dark count rate (DCR), $\gamma_\text{DCR}(I,T)$, at several temperatures are shown in Fig.\ \ref{F1}(a) on the logarithmic scale. Apart from a slight down curving at large currents, they look almost linear. The steepness in $\gamma_\text{DCR}(I)$ dependences at a fixed temperature increases with temperature. Solid lines in Fig.\ \ref{F1}(a) are the best fits with the expression for the rate of phase slips predicted by generalized two-dimensional (2D) LAMH (Langer-Ambegaokar-McCamber-Halperin) model \cite{13,14}
	\begin{subequations}
		\label{eq1}
		\begin{gather}
			\gamma_\text{DCR}(I,T)
			=
			\Omega \exp \left[
			-\frac{ \Delta F(I,T)}{k_{B}T}
			\right] ,
			\label{eq1a}
			\\
			\Delta F(I,T)
			=
			3.86 \varepsilon_\text{cond}(T) V_F(T)
			\left(1-\frac{I}{I_c(T)}\right)^{5/4} ,
			\label{eq1b}
			\\{}
			V_F(T) = \pi a^2 \xi^2(T) d ,
			\label{eq1c}
		\end{gather}
	\end{subequations}
	where $\varepsilon_\text{cond}$ is the superconducting condensation energy density at $I=0$, $\xi$ is the coherence length, $d$ is the thickness of the strip, while $a\sim1$ and the attempt rate $\Omega$ are fitting parameters. We discuss the essence of the formula and the temperature dependences of $\varepsilon_\text{cond}$ and $\xi$ in Sec.\ \ref{S3A}. From the best fits we found $a = 1.73\pm 0.14$ almost independent of temperature.
	
	At each temperature the maximum rate corresponds to the experimental critical current $I_c$. Picking up pairs of maximum rate values and corresponding currents, we plot temperature dependences of the critical current and the maximum rate $\max(\gamma_\text{DCR})$, see Fig.\ \ref{F1}(b). The maximum rate increases with the temperature as $\max(\gamma_\text{DCR}) \approx \exp(12T/T_c)$. This dependence is shown in Fig.\ \ref{F1}(b) with the straight line. The $I_c(T)$ dependence closely follows the Bardeen interpolation \cite{15} of the temperature dependence of the pair braking current $I_{B}\propto [1-(T/T_c)^2]^{3/2}$, which is also shown with the solid line in Fig.\ \ref{F1}(b).
	
\begin{figure}
		\centerline{\includegraphics[width=0.48\textwidth]{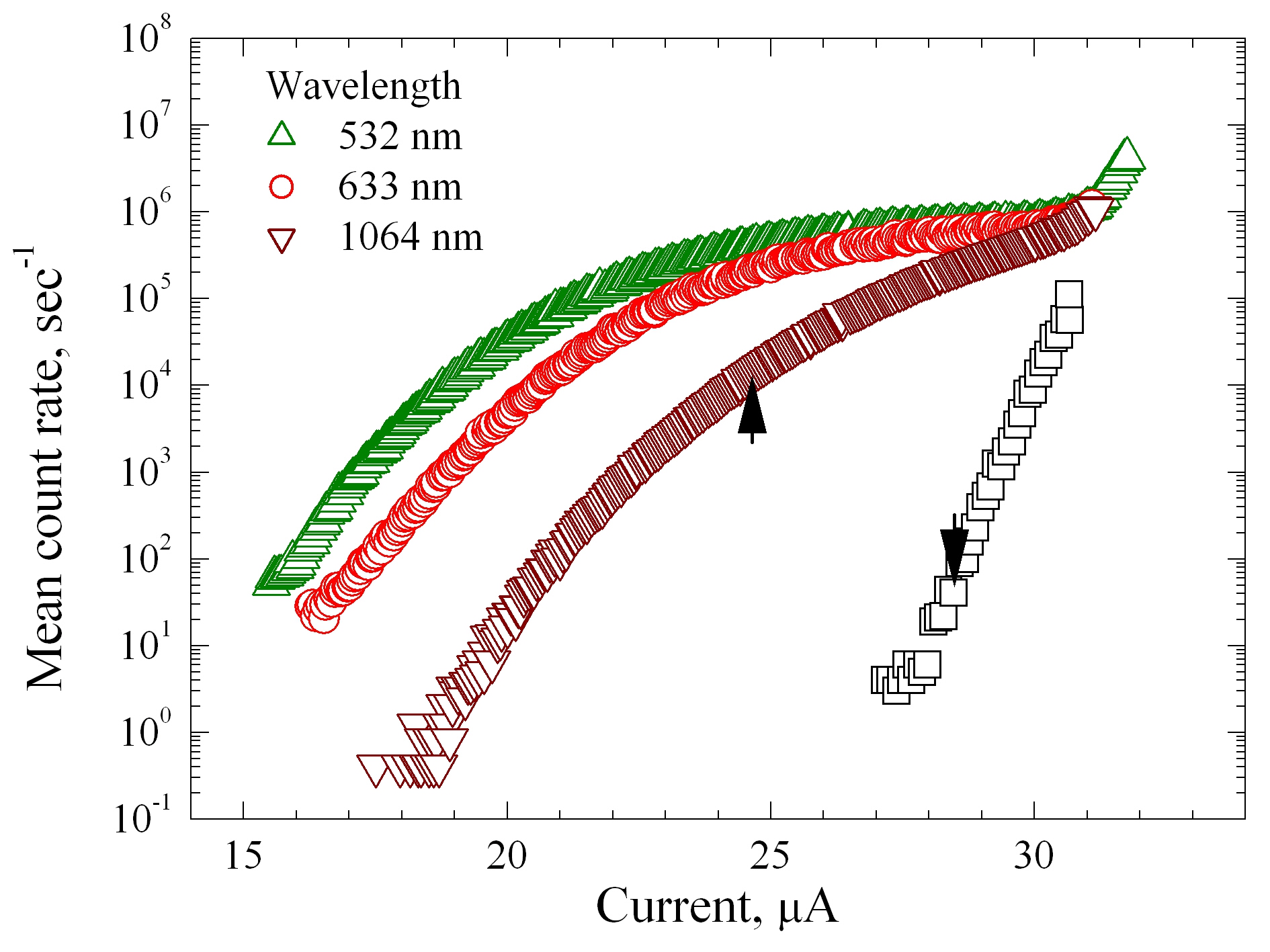}}
		\caption{DCR (squares) and PCR for different wavelengths (specified in the legend) versus current at 4.9 K. PCR at different wavelengths were scaled to the maximum value of $10^6$~sec$^{-1}$ at 31 $\mu$­A. Arrows mark the regimes for the DCR (28.4 $\mu$­A; 41 sec$^{-1}$) and PCR at 1064 nm (24.4 $\mu$­A; $1.14\times10^4$ sec$^{-1}$) for which count statistics was analyzed.}
		\label{F2}
	\end{figure}

	Current dependences of the photon count rates (PCR), $\gamma_\text{PCR}$, for three used wavelengths are shown in Fig.\ \ref{F2} along with the current dependence of the dark count rate, $\gamma_\text{DCR}$. Photon fluxes at all three wavelength were small enough to ensure that the maximum count rate $10^6$ sec$^{-1}$ is much smaller than the reciprocal dead time (approximately 10 ns). The data were acquired at 4.9 K. For photons with larger energy (550 nm wavelength) $\gamma_\text{PCR}(I)$ dependence almost saturates at $I \approx I_c$, while for low-energy photons the rate increases rapidly at all currents.

	%%%%%%%%%%%%%%%%%%%% Figure/Image No: 2 starts here %%%%%%%%%%%%%%%%%%%%
	
	%%%%%%%%%%%%%%%%%%%% Figure/Image No: 2 Ends here %%%%%%%%%%%%%%%%%%%%

	%%%%%%%%%%%%%%%%%%%% Figure/Image No: 3 starts here %%%%%%%%%%%%%%%%%%%%
	
	\begin{figure}
		\begin{center}
			\centerline{\includegraphics[width=0.48\textwidth]{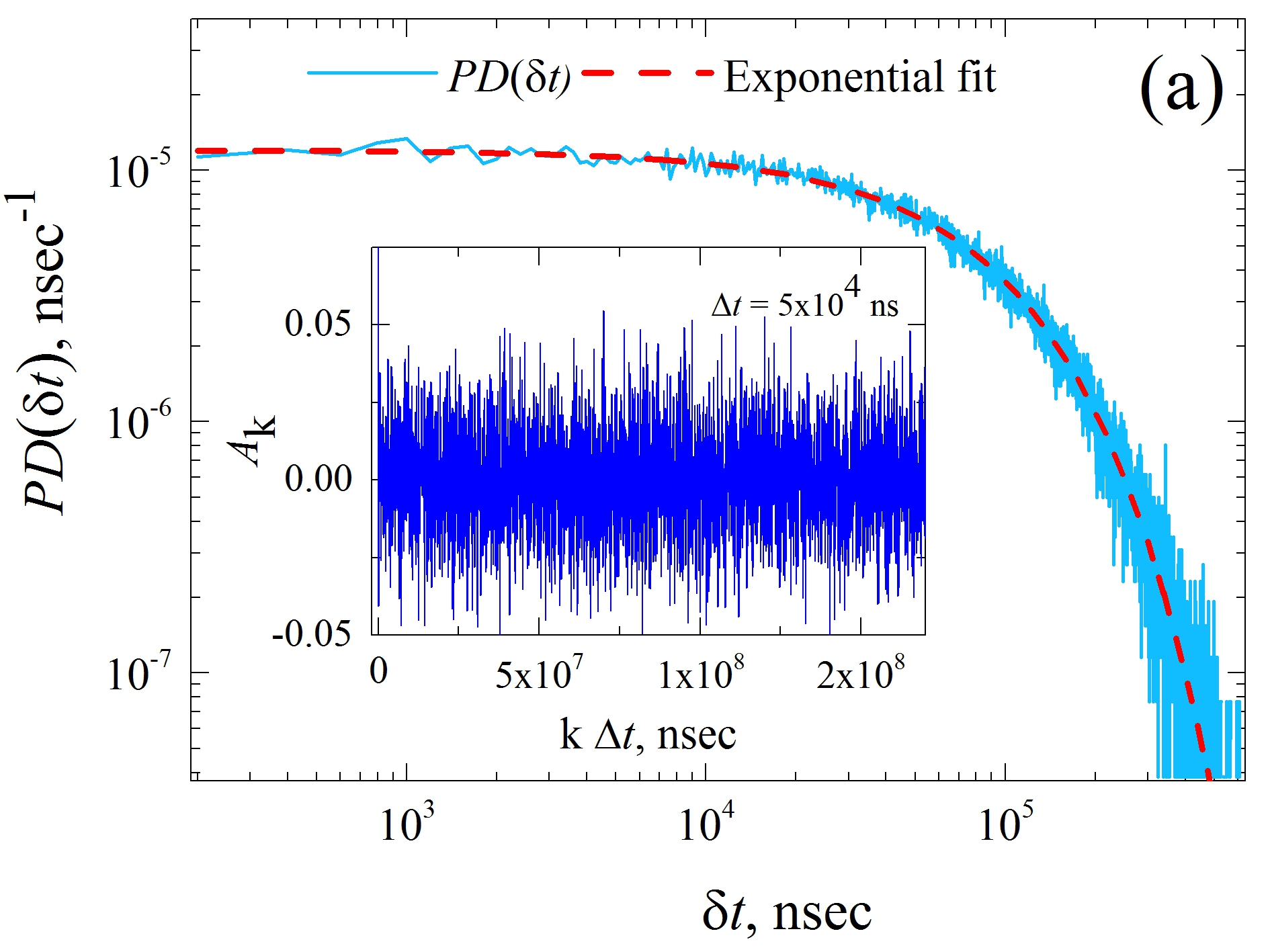}}
			\centerline{\includegraphics[width=0.48\textwidth]{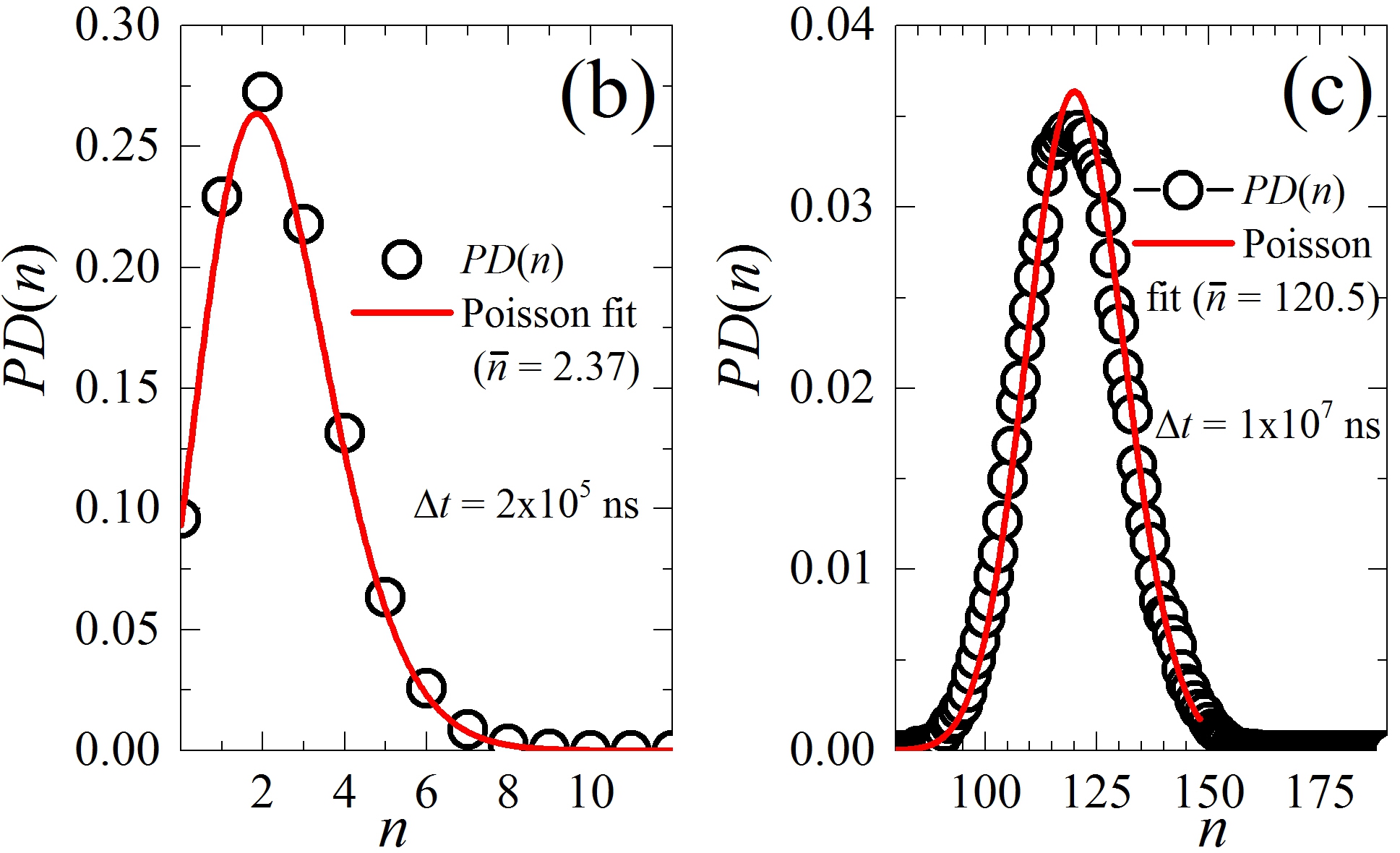}}
		\end{center}
		\caption{Statistics of photon counts for the wavelength 1064 nm. Data were acquired at 4.9 K and \textit{I} = 24.6 $\mu$­A.
			Panel (a): Probability density of the inter arrival time. Solid line presents the best exponential fit. Inset shows autocorrelation function computed according to Eq.\ (\ref{eq2}) for the time window $\Delta t = 5\times 10^4$ ns. Panels (b) and (c) show probability densities for the number of counts in time windows $2\times 10^{5}$ and $10^{7}$ ns, respectively. Solid lines present best Poisson fits. The mean values are specified in the legends.}
		\label{F3}
	\end{figure}
	
	%%%%%%%%%%%%%%%%%%%% Figure/Image No: 3 Ends here %%%%%%%%%%%%%%%%%%%%

	\subsection{Count statistics and spectra of photon counts}

	Before going to DCR statistics we prove the technique by demonstrating that photon counts delivered by the strip reproduce duly statistics of the laser light. It is known \cite{16} that arrivals of coherent photons are a discrete Poisson process for which the probability density of the time interval $\delta t$ between arrivals of two sequential photons (inter arrival time) is described by the exponential distribution $E(\delta t) = \gamma  \exp(-\gamma\, \delta t)$, where $\gamma$ is the mean arrival rate of photon. For such photon stream, the probability density for exactly $n$ photons to arrive within a given time interval $\Delta t$ is described by the binomial distribution approximated by the Poisson distribution $P(n) = (n!)^{-1} {\overline n}^{n}\exp(-\overline n)$ for small $\overline n = \gamma \, \Delta t$ and by the normal distribution $N(n) = (2\pi \overline n^{2})^{-1/2} \exp[-(n-\overline n)^2/(2\overline n)]$ for $\overline n \gg 1$.
	
	The panel (a) in Fig.\ \ref{F3} shows the normalized probability density, $PD(\delta t)$, of the time interval between two adjacent photon counts. The solid line is the best fit with the function $f(\delta t) = \gamma \exp( - \gamma\, \delta t)$, where $\gamma = 1.2\times 10^4$ sec$^{-1}$ is very close to the measured mean photon count rate of photons $\gamma_\text{PCR} = 1.14 \times 10^4$ sec$^{-1}$ (Fig.\ \ref{F2}). It is clearly seen that the measured probability density follows exactly the form expected for the Poisson process.
	
	To cross check that the process we deal with is indeed Poisson, we measured occurrence times $t_{i}$ of all count events within an acquisition time of a few minutes. We then converted the array $\{t_{i}\}$ of measured arrival times into the array of the numbers of events $\{n_j\}$ which occurred in a given time interval $\Delta t$. We put intervals on the array $\{t_{i}\}$ in such a way that each event $t_{i}$ starts its own interval until the end of the interval hits the end of the massive. We then computed probability densities, $PD(n)$, for exactly $n$ events to occur within the interval for two vastly different values of this interval $2\times10^{5}$ and $10^{7}$ ns. The results are shown in the panels (b) and (c) in Fig.\ \ref{F3} with the best Poisson fits obtained for mean values $\overline n = 2.37$ and 120.5, correspondingly. Using the identity $\overline n = \gamma \Delta t$, we found the best fit values $\gamma = 1.19\times 10^{4}$ sec$^{-1}$ and $1.21\times10^{4}$ sec$^{-1}$, which are very close to the experimentally measured mean count rate $\gamma_\text{PCR}=1.14 \times 10^{4}$ sec$^{-1}$.
	
	Finally, we computed the autocorrelation function for the variable $n$. To save computation time, we reduce the number of intervals  \(  \Delta t \)  by splitting the total acquisition time of the array $\{t_{i}\}$ in equal adjacent intervals and count the number of events $n_{j}$  in each of them. The second order autocorrelation function is defined as
	\begin{equation}
		A_{k}
		=
		\frac{\sum _{j=1}^{N/2} (n_{j}-\overline n) (n_{j+k}-\overline n)}
		{\sum _{j=1}^{N/2} (n_{j}-\overline n)^2} ,
		\label{eq2}
	\end{equation}
	where $N$ is the number of elements in $\{n_{j}\}$ and $\overline n=\gamma_\text{PCR} \Delta t$. The result obtained for $\Delta t = 2\times 10^{5}$ ns is shown in Fig.\ \ref{F3}(a). The autocorrelation function is centered at zero for all $k$ that evidences statistical independence of photon counts for the whole array at the time scale larger than $\Delta t$. We showed exemplarily data acquired for the wavelength 1064 nm at 5 K and $I = 24.6$~$\mu$­A. Qualitatively same results were obtained at all other wavelengths and currents presented in Fig.\ \ref{F2} apart from the region $I \approx I_c$, where rates of photon and dark counts are comparable. Hence the strip as a single photon detector duly reproduces statistics of coherent photons. This justifies the technique we choose. By recording Poisson statistics of the number of photon counts we also proved that the incandescent light of the monochromator contained large enough number of modes to ensure Poisson statistics of photons \cite{16}. We shall note here that measuring inter arrival statistics for photon counts from continuous coherent light source (laser) is a quick and simple method for qualifying single photon detectors as compared to techniques relying on pulsed lasers and two detectors after a beam splitter.
	
	%%%%%%%%%%%%%%%%%%%% Figure/Image No: 4 starts here %%%%%%%%%%%%%%%%%%%%
	
	\begin{figure}
		\centerline{\includegraphics[width=0.48\textwidth]{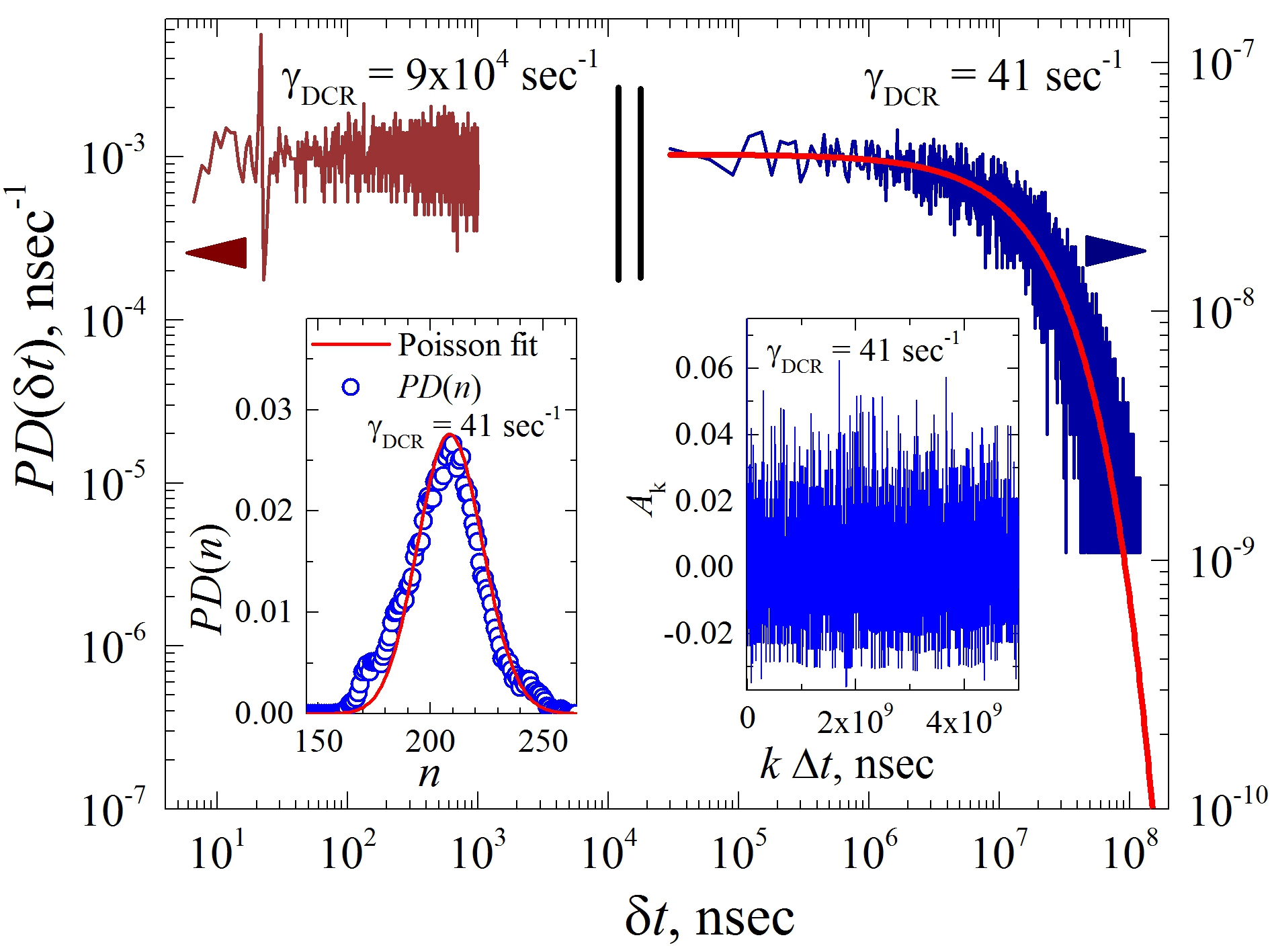}}
		\caption{Statistics of dark counts. Main panel: probability density of inter arrival times for two different mean count rates. Solid line is the best exponential fit to the data obtained at the smaller rate. Left inset: probability density of the number of counts in the time window 5 sec at the smaller mean count rate. Solid line shows Poisson distribution with $\overline{n} = 209$. Right inset: Autocorrelation function for the smaller count rate.}
		\label{F4}
	\end{figure}
	
	%%%%%%%%%%%%%%%%%%%% Figure/Image No: 4 Ends here %%%%%%%%%%%%%%%%%%%%

We further apply the technique described above to evaluate statistics of dark counts. The results shown in Fig.\ \ref{F4} were acquired at 4.9 K and two currents $I = 28.4$ $\mu$­A and 30.7 $\mu$­A. The corresponding mean dark count rates were 41 sec$^{-1}$ and $9\times10^{4}$ sec$^{-1}$. The main panel shows probability density of inter arrival time of dark counts for these two currents. At the smaller current (right axis) probability density decreases exponentially as expected for non-correlated Poisson process. Solid line represents the best exponential fit with $\gamma = 43$ sec$^{-1}$, which is almost equal to the experimental value $\gamma_\text{DCR} = 41$ sec$^{-1}$ (Fig.\ \ref{F2}). Probability density for small intervals of the order of a few nanoseconds could be obtained only for much larger mean count rates corresponding to larger currents. The curve affiliated with the left axis was computed from the data acquired at the larger current. It shows the beginning of the drop at small intervals which we attribute to the recovery of the strip after the count event. Full recovery occurs after approximately 10 ns that is twenty times the full width at half maximum of the voltage transient after the amplifier. The spike at 20 ns is due to the reflection of the voltage pulse at the input of the amplifier. A small bump in the curve that appears right after the recovery has been completed may signal the presence of after pulsing. Inserts in Fig.\ \ref{F4} show probability density of the number of counts in the time window 5 sec (left) and the autocorrelation function computed for the time window 1 msec (right). While the autocorrelation function evenly scattered around zero thus confirming that the process is not correlated for times larger than 1 millisecond, the probability density for the number of counts deviates from the Poisson distribution. Solid line shows Poisson distribution for the mean value $\overline n = 209$ which provides maximum in the experimental \textit{PD}(\textit{n}) dependence. The corresponding standard deviation $\sigma = \sqrt{\overline n}$ equals 14.5, while numerically computed for experimental data points standard deviation 16.8 is slightly larger.  We have to note that at count rates in excess to 50 sec$^{-1}$ the statistics of dark counts drastically changes. The autocorrelation function reveals periodical oscillations, $PD(n)$ broadens with respect to the Poisson distribution with the same mean value and the probability density of the inter interval time rolls off as $\exp[-(\gamma_\text{DCR} \delta t)^p]$ with $p\approx 0.85$.  We believe that this is an artificial effect created by the stabilizing feedback in the electronics.
	
	%%%%%%%%%%%%%%%%%%%% Figure/Image No: 5 starts here %%%%%%%%%%%%%%%%%%%%
	
	\begin{figure}
		\centerline{\includegraphics[width=0.48\textwidth]{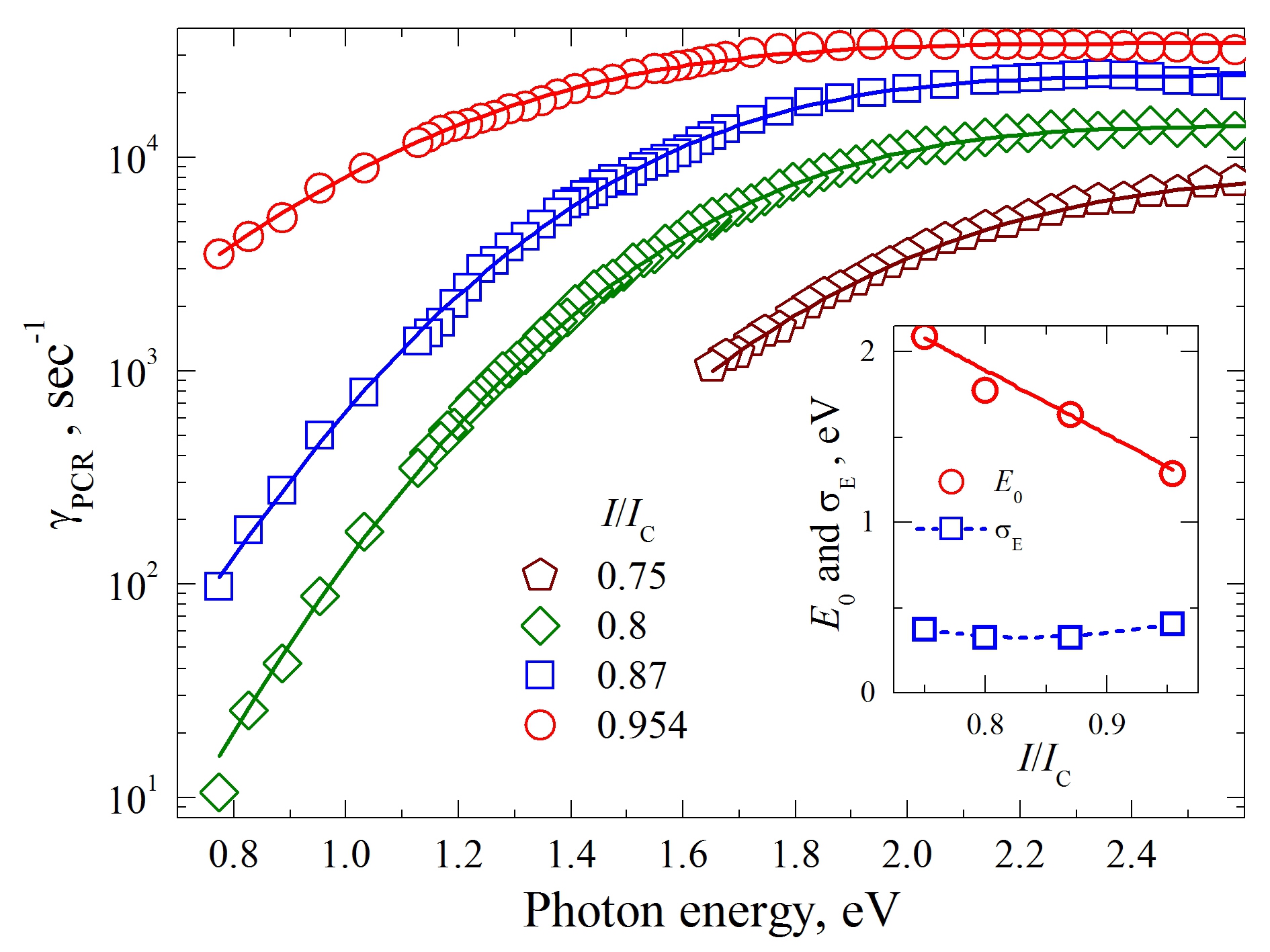}}
		\caption{Spectra of the mean photon count rate at different currents. Solid lines are best fits with Eq.\ \eqref{eq-PCR}. Inset shows the best fit values for the cut-off energy $E_0$ and the width of the transition $\sigma_{E}$. Solid line in the inset shows the model fit, and dashed line is guide to the eyes.}
		\label{F5}
	\end{figure}
	
	%%%%%%%%%%%%%%%%%%%% Figure/Image No: 5 Ends here %%%%%%%%%%%%%%%%%%%%
	
	Figure \ref{F5} shows spectra of photon count rates at different currents obtained at a temperature of 4.5 K. Measured count rates were normalized to the fixed photon flux at the specimen. There is a continuous increase of the count rate at all\ currents. At relative currents larger than 0.8, $\gamma_\text{PCR}$ saturates in the shown range of photon energies. The saturation level increases with the current. Solid lines represent fits by the formula
\be
  \gamma_\text{PCR}(E)
  \propto
  1+\mathop{\rm erf}
  \left(
    \frac{E-E_0}{\sqrt{2}\, \sigma_E}
  \right) ,
\label{eq-PCR}
\ee	
which we discuss below in Sec.\ \ref{S3B} (erf is the error function). The best fit values $E_{0}$ and $\sigma_{E}$ shown in the inset in Fig.\ \ref{F5} formally define the effective cut-off energy in the spectrum of $\gamma_\text{PCR}$ and the width of the transition from zero efficiency to the saturated value.

\section{Fluctuation analysis}
	\label{S3}

	In this section we analyze different types of local fluctuations in a current-carrying superconducting strip. We assume that each photon count or dark count is an intrinsically deterministic event. Whether the count occurs depends on the instantaneous local value of the internal electron energy in a relevant nucleation volume located somewhere in the superconducting strip. For a dark count to occur, thermal fluctuations in respective nucleation volume have to drive the superconducting system out of the equilibrium state which is surrounded by an activation barrier in the free energy. Once this happens superconductivity is destroyed locally that deterministically triggers a count event. As the reference point for the Helmholz free energy of electrons, $F$, we take its local minimum value, $F_{0}$, corresponding to the equilibrium uniform superconducting current state. At any finite temperature $F$ is the subject to local thermal fluctuations. The activation barrier to overcome is $\Delta F = F_{+}-F_{0}$, where $F_{+}$ is the energy of the saddle point configuration. The probability density of the interarrival time and, consequently, the dark count rate are proportional to $\exp [-(F-F_0)/(k_{B}T)]$ as given by the Boltzmann distribution.
	If the activation barrier is non-uniform over the area of the strip, dark counts presumably occur in the areas with the smallest $\Delta F$. Such areas are defined by constrictions \cite{Kermann 2007} and static local fluctuations in the strip thickness \cite{Cheng 2017} or in the superconducting energy gap \cite{Rod 2020}. The barrier also depends on the position across the strip \cite{7, 10}. The presence of such fluctuation does not affect the statistics of dark counts but modifies their mean count rate.	
	
	Differently to dark counts, a photon count comes from the hot-spot created by the photon around absorption site.  Since absorption sites are evenly distributed over the strip surface, local fluctuations of any kind in the activation barrier smear deterministic, step-like dependence of the mean photon count rate $\gamma_\text{PCR}(E)$ on the photon energy $E$. As it is true for dark counts, fluctuations in the barrier do not affect the statistics of photon counts for fixed photon energy. Besides factors mentioned above, the activation barrier is affected by local variations of the electron enthalpy which in our case is identical to the internal electron energy. One of the sources of local variations in the internal energy is Fano fluctuations [8] which randomize the portion of the photon energy delivered to electrons within the hot-spot. Another source, which we introduce below in Section II B, is local (non-uniform) thermal fluctuations of the internal electron energy in small cells spread over the strip. Phonons in the strip itself together with phonons in the substrate serves for each such cell as thermal bath. Fluctuations in different cells are statistically independent unless their sizes are smaller than the correlation length of fluctuations. Statistical equilibrium thermodynamics \cite{Landau} predicts the variance of such fluctuations proportional to the mean squared temperature and the heat capacity of the cell. Uniform microscopic model of such fluctuations  \cite{U_FLUCT} based on Langevin equations and random heat exchange between the whole strip and the underlying substrate describes quantitatively noise spectrum that was observed in multiphoton photodetectors such as transition edge sensors \cite{TES}, hot-electron bolometers \cite{HEB}, and microbolometers \cite{MBL}. In these detectors, however, local fluctuations on any scale smaller than the strip size are averaged out and cannot be detected experimentally.

	\subsection{Dark count rate}
	\label{S3A}
	
\subsubsection{\textbf{Generalized LAMH theory}}

	In the case of dark counts, the above phenomenological picture of fluctuations in the free energy can be put on firm theoretical grounds as the problem belongs to the class of thermal decay of a metastable state in equilibrium statistical mechanics. The threshold value $\Delta F$ as well as the fluctuation volume $V_{F}$ is then determined by an optimal fluctuation that provides the saddle point rather than the minimum to the free energy. The theory of thermally assisted phase slips in \textit{quasi-one-dimensional} (1D) superconducting wires [with the cross-section $S \ll \xi^2(T)$] was developed in Refs.\ \cite{13,14} (LAMH theory).
	Working within the Ginzburg-Landau approximation applicable in the vicinity of $T_c$, it is possible to identify the saddle-point (instanton) configuration of the order parameter field that should be reached by thermal fluctuations. Once it is approached, further evolution of the overheated region is not-probabilistic leading eventually to the relaxation of phase winding and appearance of a detectable voltage pulse.
	Hence the expression for the decay rate takes the form of Eq.\ \eqref{eq1a}, with the activation barrier given by Eq.\ \eqref{eq1b} (see Appendix).
	In Eq.\ \eqref{eq1b}, $\varepsilon_\text{cond}(T)$ is the condensation energy density in the absence of supercurrent, the last factor accounts for the flattening of the barrier as $I\to I_c$, and $V_F(T) = S \xi(T)$ is the effective volume of the fluctuation region in a 1D wire [here is the difference with the 2D Eq.\ \eqref{eq1c}!].
	
	The original LAMH theory describes thermal decay of the supercurrent state in 1D wires near $T_c$. In order to generalize it to arbitrary temperatures below $T_c$ one has to go beyond the GL approximation and to use the full set of equations for dirty superconductors, including the Usadel equation for quasiparticles and the self-consistency equation for the order parameter. This program was realized in Ref.\ \cite{19,Marychev2017}, where the function $\Delta F(I,T)$ was obtained numerically for arbitrary currents and temperatures. With rather good accuracy it can still be cast in the form (\ref{eq1b}) with $V_F(T) = S \xi(T)$, where one should properly adjust the GL expressions for $\varepsilon_\text{cond}(T)$ and  $\xi(T)$ at low temperatures \cite{Marychev2017}. For practical purposes we propose the following analytical approximations valid for the whole temperature range (see Appendix A):
	\begin{gather}
		\varepsilon_\text{cond}(T) = 1.556 \, N_0 (k_{B}T_c)^2 (1-0.132\hat t^{2})^2 (1-\hat t^2) ^2 ,
		\nonumber
		\\
		\xi(T) = 0.667 \coth \left( 0.655 \, \hat t^{-1}\sqrt{1-\hat t^{2.64}} \right) \sqrt{\frac{\hbar D}{k_{B}T_c}} ,
		\label{eq4}
	\end{gather}
	where $\hat t=T/T_c$ is the reduced temperature, $N_{0}$ is the density of states at Fermi level per one spin projection and \textit{D} is the diffusion coefficient in the normal state.
	
In the 2D geometry, an accurate theory of the thermally assisted phase slips is missing. While it is generally accepted that dark count events in superconducting strips are triggered by vortex (or vortex-antivortex pair) motion \cite{Vodolazov2012}, the value of the activation barrier for vortex unbinding remains to be determined. In the simplest approximation, when the vortex is considered as a zero-size particle, the problem was addressed in Ref.\ \cite{20}. From the theoretical side, the drawback of this analysis is that the distance of the vortex from the strip edge (or the vortex-antivortex separation) is of the order of  $\xi(T)$, except for extremely small currents $I \ll I_c$. Hence the contribution of the vortex core is essential and cannot be neglected. From the experimental side, the vortex hopping model \cite{20} does not reproduce the current dependences at all temperatures with a fixed set of fit parameters. Moreover, the temperature dependence of the preexponent factor $\Omega$ is much steeper than the experiment shows.
	
	In order to describe our experimental data on dark count rates in 2D, we propose a phenomenological model based on (i) LAMH theory in 1D, (ii) its generalization to lower temperatures \cite{19,Marychev2017} as summarized by Eq.\ (\ref{eq4}), and (iii) the observation that at $I \sim I_c$ the relevant spatial scale in the vortex model is of the order of $\xi(T)$ \cite{Vodolazov2012,20}. These arguments suggest that the activation barrier in 2D can be written in the same form (\ref{eq1b}) but with the fluctuation volume $V_F(T) \sim  \xi^2(T) d$. We refer to this ansatz as the generalized 2D LAMH model.
	
	The set of Eqs.\ (\ref{eq1}) with $D = 5\times10^{-5}$ m$^2$ s$^{-1}$, $N_0 = 1.5\times 10^{47}$ J$^{-1}$ m$^{-3}$ \cite{21} and $d = 5$ nm were used to fit experimental data shown in Fig.\ \ref{F1}a. From the best fit we found that the effective fluctuation volume corresponds to $V_{F}(T) = \pi a^2 \xi^2(T) d$ with $a \approx 1.73$.

\subsubsection{\textbf{Stochastic real-time description}}
\label{SSS:dark-stoch}

The activation exponent for the dark-count rate in Eq. \eqref{eq1} was obtained in a purely thermodynamic way as a free-energy cost $\Delta F$ of an optimal fluctuation at a given temperature $T$. Such an approach leaves open the question of the development of fluctuations in real time, which is important  for the analysis of photon counts. To describe this process, one has to deal with temporal fluctuations of the local density of the internal electron energy, $u(\mathbf{r},t)$. These thermal fluctuations are inherent to equilibrium systems and are caused by coupling to the thermostat, whatever the microscopic mechanism of this coupling is.
Qualitatively, development of the optimal fluctuation in time can be described in terms of the internal electron energy in the fluctuation volume $V_U$ given by $U(t)=\int u(\mathbf{r},t)\,dV_U$. Its fluctuations around the mean $\overline U$ are described by the normal probability distribution function
\begin{equation}
  f(U)
  =
  \frac{1}{\sqrt{2\pi}\sigma_{U}}
  \exp \left[ -\frac{1}{2 \sigma_U^2} \left( U-\overline U \right)^2 \right],
\label{eq5}
\end{equation}
with the variance $\sigma_U^2=k_B T^2 cV_U$ being determined by the specific heat capacity of electrons, $c$, according to the theory of thermodynamic fluctuations \cite{Landau}.

Time-dependent fluctuations of $U(t)$ can be considered as a continuous stochastic Ornstein-Uhlenbeck (OU) processes \cite{22} in statistically independent volumes spread over the strip. Each count temporarily reduces the current through the strip and resets OU processes in all volumes. Hence the question when the next count may first occur reduces to the problem of the first hitting time in the OU process, i.e., the time which elapses from the start of the process until the electron energy $U$ for the first time hits the preset boundary $\overline U + \Delta U$.
Although the probability density of the first hitting time can be computed exactly \cite{23}, we need only the asymptotic for times much larger than the life time of the fluctuation $\tau_F$.
This asymptotic is an exponential function of the form $g \exp(-g \delta t)$ \cite{24,25}. For $\Delta U \gg \sigma_U$, the theory of OU processes predicts $g=\Delta U \exp[ - \Delta U^{2}/(2\sigma_U^2)]/(\sqrt{2\pi}\sigma_U \tau_F)$.
Obviously, $g$ represents the rate of dark counts in terms of the stochastic process.
Hence, one should identify $(\Delta U)^2/2\sigma_U^2$ with $\Delta F/k_BT$, and $\Delta U/(\sqrt{2\pi}\sigma_U \tau_F)$ with the attempt frequency $\Omega$.

	\subsection{Photon count rate}
	\label{S3B}

\subsubsection{\textbf{Previously discussed broadening mechanisms}}
	
Our spectra of the detection efficiency were fitted by Eq.\ (\ref{eq-PCR}). The best fits to experimental dependences $\gamma_\text{PCR}(E)$ are shown with solid lines in Fig.\ \ref{F5}. The inset represents the fitting parameters $E_0$ and $\sigma_E$ for different bias currents.
We see that $\sigma_E\approx330$ meV, nearly independent of $I/I_c$.

Below we discuss several mechanisms that might be responsible for the spectral broadening and show that all of them underestimate the magnitude of the effect. As mentioned above, there are several mechanisms, which may contribute to the spectral width, $\sigma_E$, of the dependence $\gamma_\text{PCR}(E)$:
\begin{itemize}
\item
Dependence of the detection current on the position of the absorption site across the strip was theoretically studied in Refs.\ \cite{7,10}. The difference between the maximum and the minimum of the local detection current within the strip width along with the dependence of the maximum detection current on the photon energy give an estimate $\sigma_E \approx 50$ meV, which is much smaller than our experimental values $\sigma_E \approx 330$ meV.

\item
Fano fluctuations mediated in thin films by the escape of high-energy phonons would result in $\sigma_E=(G \zeta h E \Delta)^{1/2}$, where the effective Fano factor $G \approx 1$ \cite{9} and the upper boundary for the quantum efficiency
$\zeta$ = 0.5 as defined by the ratio of electron and phonon heat capacities \cite{Sidor 2020}. With the energy gap $\Delta = 2.05 \, k_B T_c$ and the photon energy 1 eV, we obtain $\sigma_E$ = 33 meV, which is an order of magnitude smaller than our experimental values.
Furthermore, Eq.~{(\ref{eq5})} with $\sigma_E \propto \sqrt{E}$ poorly fits our experimental spectra.

\item
Effect of the film nonuniformity is hard to address quantitatively. High resolution transmission electron microscopy of our films \cite{21} revealed granules with the size of the order of the film thickness. Granulas with the dominant size even less than the film thickness were also found in thicker NbN films prepared by a different technology \cite{LosAlam}.
Typically, the size of granules is distributed according to either log-normal \cite{32} or inverse cubic function \cite{LosAlam} with a standard deviation not larger than the film thickness. Therefore, for our films with the thickness equal to the coherence length, we do not expect a noticeable effect of the film granularity on the PCR spectra.
Inspection of our films with an atomic force microscope (Ref.\ \cite{21} and Appendix B) returned the correlation radius in the film thickness  $ R_d <$ 10 nm and the standard deviation on a larger length scale $\sigma_d <$ 0.13 nm. Hence, expected contribution of the thickness fluctuations to the relative spectral broadening $\sigma_d / d =$ 0.026 ($d =$ 5 nm)
is much less than the relative  experimental spectral broadening $\sigma_E / E_0 >$ 0.2.

\item
Non-uniformity of the energy gap in 2.1 nm thick NbN films was studied with a scanning-tunneling microscope (STM) \cite{Rod 2020}. For the gap variations the authors found a correlation radius  of $R_\Delta \approx 28$ nm and the relative gap fluctuations $\sigma_\Delta / \Delta \approx$ 0.08. This is still a few times less than the relative spectral broadening found experimentally.
\end{itemize}

\subsubsection{\textbf{Local thermal fluctuations}}

Facing the finding that the known fluctuation mechanisms and non-uniformities do not quantitatively explain spectral broadening of the photon count rate, we invoke \emph{local thermal fluctuations} on the length scale $l_U$ corresponding to the effective fluctuation volume $V_U=\pi l_U^2d$ as the leading source of broadening. In the absence of photons such fluctuations have been discussed in Sec.\ \ref{SSS:dark-stoch}, and now we calculate their effect on the spectral broadening $\sigma_E$.

Upon arrival of a photon, a part of its energy released to the electronic system increases the internal energy of the latter within a hot spot with the radius $l_\text{hs}$. Assuming that the delivered energy and the internal electron energy are instantaneously additive, one can qualitatively describe the process of photon absorption as a sudden change of the internal energy $U\to U+\varsigma^* E$ in the effective volume $V_U$. Here $E$ is the photon energy, $\varsigma^*=\zeta (l_U / \l_\text{hs})^2$ is its fraction that heats electrons in the fluctuation volume $V_U$, and $\zeta$ is the quantum efficiency (the portion of the photon energy delivered to electrons in the hot spot).
This approach is meaningful only when $l_U\leq l_\text{hs}$. For $l_U\gg l_\text{hs}$, part of the photon energy delivered to the hot spot does not noticeably change the electron energy in the effective fluctuation volume $V_U$.

The photon is detected if the elevated energy $U+\varsigma^* E$ exceeds $\overline U+\Delta U$, where $\Delta U$ is the relevant activation barrier. Since $U$ itself is a random quantity with the distribution \eqref{eq5}, the sharp deterministic dependence of the photon count rate on the photon energy becomes smeared:
\be
		\gamma_\text{PCR}(E)
      \propto
		\int_{\overline U+\Delta U-\varsigma^* E}^{\infty}f(U) dU
,
		\label{eq6}
\ee
that leads to Eq.\ \eqref{eq-PCR}, with $E_{0}= \Delta U / \varsigma^*$ being the cut-off photon energy in the deterministic detection scenario, and the following relation between the spectral width $\sigma_E$ and the standard deviation $\sigma_U$:
\be
  \sigma_E 
  = \frac{\sigma_U}{\varsigma^*} 
  = \sigma_U \sqrt{\frac{E_0}{\Delta U \varsigma^*}}
  .
  \label{eq7}
\ee

Using the measured spectral width of $\gamma_\text{PCR}(E)$, we can estimate the effective size of a relevant fluctuation, $l_U$, and check the concept of local thermal fluctuations developed above for consistency. Macroscopically, the increase $\Delta U$ in the mean energy, which suppresses superconductivity in the current carrying state, corresponds to the steady-state change, $\Delta T$, in the ambient temperature that reduces the value of the critical current to the bias current, i.e. satisfies the condition $I=I_c(T+\Delta T)$. Using analytical fit to experimental $I_c(T)$ dependence (Fig.\ \ref{F1}b), we computed $\Delta T(I,T)$ for the temperature and currents at which the spectral data (Fig.\ \ref{F5}) were acquired.
For the best linear fit in the inset of Fig.\ \ref{F5} we obtain $E_0/\Delta T = 0.72$ eV/K. Defining the threshold in the electron energy as $\Delta U = c V_U \Delta T$ and recalling that $E_0 = \Delta U/\varsigma^*$, we see that the factor $c V_U$ cancels out from the last equality in Eq.\ \eqref{eq7} that allows defining the fraction $\varsigma^*$ from experimental data without knowing $V_U$ as $\varsigma^* = (k_B T^2/\sigma_E^2)(E_0/\Delta T)$. Substituting $E_0/\Delta T = 0.72$, $\sigma_{E} = 330$ meV and $\zeta = 0.5$, we find $\varsigma^*= 0.017$ and thus establish a relation between the size of a relevant fluctuation and the size of the hot spot: $l_U\approx 0.2\: l_\text{hs}$.

The radius of the hot spot created by a photon with the energy of 1 eV is a few tens of nanometers \cite{27}. It can be estimated as $l_\text{hs}\approx (D \tau_\text{hs})^{1/2} =$ 45 nm, where $\tau_\text{hs} \approx 40$ ps is the life time of the hot spot \cite{28}. This gives an estimate $l_U \approx 9$ nm. Differently, using the heat capacity of electrons in the normal state $c=2\pi^2k_B^2N_0T_c/3$ and the values of $E_0/\Delta T$ and $\varsigma^*$, we find $l_U \approx 12$ nm. The two estimates are very close. However, due to uncertainty in the electron heat capacity in the current carrying superconducting state \cite{29} the second estimate for $l_U$ is less reliable. We would like to note here that the estimated size of the effective fluctuation volume  for photon counts  is very close to the size of the nucleation volume for dark counts $a \xi\approx$ 8.5 nm determined in Sec.\ref {S3A} ($\xi = 5$ nm for our operation temperature, Eq. \ref {eq4}). This closeness might indirectly evidence that in both cases we deal with the same fluctuation phenomenon.

	\section{Discussion and conclusion}
	\label{S4}
	
In this paper, we demonstrate that the phenomenological model of local thermal fluctuations describes quantitatively our experimental data for the spectral broadening of the photon detection efficiency with the reasonable value of the effective fluctuation volume which almost coincides with the nucleation volume of dark counts. We supposed that local thermal fluctuations are statistically independent from the amount of energy delivered by the photon into the fluctuating cell. This may be not strictly correct if correlation time of the local fluctuations is much smaller than the life time of the hot spot. The fluctuation model itself is valid unless the correlation length, $R_U$, of thermal fluctuations exceeds $l_U$.
The estimate $R_U\approx (D\tau_\text{EP})^{1/2} =$ 5 nm, where $\tau_\text{EP}\approx$ 0.6 ps is the relaxation time of the electron energy via phonons in the strip at the transition temperature averaged over electron distribution \cite{Sidor 2020}, gives  $R_U < l_U$. 
Although random heat exchange between electrons in the strip and the substrate via phonons will further modify the correlation length, the effect of the substrate in our case is reduced by the almost ballistic transport of phonons in the film on the lenght scale of $R_U$. Indeed, the phonon mean free path in the film $l_{ph} = u_{ph} \tau_{PE}\approx$ 1.5 nm is of the order of the estimated correlation length. Here $u_{ph} =$ 2.5 m sec$^{-1}$ is the phonon velocity and $\tau_\text{PE} \approx \tau_\text{EP} =$ 0.6 ps is the relaxation time of the phonon energy via phonon-electron interaction in our films \cite{Sidor 2020}. Bearing in mind possible effect of the substrate, we rather consider the estimated value of the correlation length as the lover boundary for the actual value. 
Rigorous microscopic theory of local thermal fluctuations should relax these shortcomings. However, such theory remains beyond the scope of this work.

	We would like to note here that the width of the $\gamma_\text{PCR}$ spectra, which we obtained experimentally, corresponds to the ultimate value of the timing jitter in photon detection by superconducting strips. Indeed, with the mean slope of the count delay time (latency) vs photon energy of $12\hbar/(k_BT_c)$ per one electron-volt (Fig.\ 5b of Ref.\ \cite{7}) and $\sigma_E \approx 0.33$ eV 
(Fig.\ \ref{F5} of the present text) we arrive at the jitter (standard deviation) of approximately 2.7 ps that is close to ultimate reported values \cite{4,5}.
	
	In conclusion, we have shown that thermal fluctuations in the free energy, which drive the cell with the size of a few coherence lengths into the normal state over the saddle point in the potential well, quantitatively describe temperature and current dependences of the dark count rate in a current carrying superconducting strip. We have furthermore shown that equilibrium thermal fluctuations in a cell of approximately the same size, explain qualitatively the broadening of the deterministic spectral cut off in the detection efficiency.

	\acknowledgments

	A.D.S. acknowledges support by German ministry of Science and Education (BMBF) in the framework of the ERA.Net RUS Plus program (Project TreSpRaE, ID 88). M.S. acknowledges support by the Helmholtz Research School on Security Technologies. M.A.S. acknowledges support from the Russian Science Foundation (Project No.~20-12-00361). Authors are greatly thankful to Arkady Pikovsky and Vadim Kaushansky for highlighting the connection between the problem of the first hitting time in stochastic continuous processes and the rate of discrete counts, and to Anton Andreev and Mikhail Feigel'man for stimulating discussions.

	\appendix

	\section{LAMH theory and beyond}
	\label{A}
	\subsection{Activation barrier in the LAMH theory}
	
	The LAMH theory \cite{13,14} applicable in the vicinity of $T_c$ gives the following implicit expression for the activation energy in the 1D case [Eq.\ (2.13) of Ref.\ \cite{14}]:
	\be
	\label{FLAMH}
	\Delta F(I,T)
	=
	\eps_\text{cond}(T) S \xi(T) U(I) ,
	\ee
	where $\eps_\text{cond}(T) = H_c^2(T)/8\pi = 4\pi^2N_0k_B^2(T_c-T)^2/7\zeta(3)$
	is the condensation energy density in the absence of supercurrent in the GL region, $S$ is the wire cross section, and $\xi_\text{GL}(T)$ is the GL coherence length.
	The current dependence of the barrier is captured by the last term in Eq.\ \eqref{FLAMH} given by
	\be
	U(I)
	=
	\frac{2^{7/2}}{3} \sqrt{1-3 \kappa ^2}-8 \kappa
	\left(1-\kappa ^2\right) \arctan\frac{\sqrt{1-3 \kappa ^2}}{\sqrt{2}
		\kappa} ,
	\ee
	where the dimensionless parameter $\kappa$ is related to the current by means of
	\be
	I = 8\pi \eps_\text{cond}(T) S \xi(T) \frac{c}{\Phi_0} J
	, \qquad
	J = \kappa(1-\kappa^2) .
	\ee
	The critical current corresponds to
	$\kappa_c = 1/\sqrt{3}$ and $J_c = 2/3\sqrt{3}$.
	
	The function $U(I)$ vanishing at $J=J_c$ can be expanded at $J\to J_c$ as
	\be
	\label{U45}
	U(I)
	\approx
	\frac{64}{15} \frac{2^{1/4}}{3^{1/4}}
	\left(1-\frac{I}{I_c(T)}\right)^{5/4} .
	\ee
	Remarkably, Eq.\ \eqref{U45} appears to be a very good approximation for the function $U(I)$ for all currents, with a few percent discrepancy even at $I=0$. For operational purposes we will replace $U(I)$ by the 5/4-power approximation \eqref{U45}. Then isolating the factor $(64/15) (2/3)^{1/4} = 3.86$, we arrive at Eq.\ \eqref{eq1b} with $V_S(T)=S\xi(T)$.
	
The LAMH theory applies at $T\to T_c$. Its generalization to arbitrary temperatures has been done in Ref.\ \cite{19,Marychev2017}, where the activation barrier was obtained by numerical solution of the Usadel and self-consistency equations. It can be reasonably approximated by the LAMH expression \eqref{FLAMH}, provided one uses an intelligent generalization of $\eps_\text{cond}(T)$ and $\xi(T)$ to arbitrary $T$.
	
Below we calculate $\eps_\text{cond}(T)$ and propose a natural generalization of $\xi(T)$ to arbitrary temperatures.
An alternative approach would be to use the relation $\eps_\text{cond}(T)\xi(T) \sim \hbar j_\text{dep}(T)/e$ exploited in Ref.~\cite{Marychev2017} and fit the temperature dependence of the depairing current density $j_\text{dep}(T)$.

	\subsection{Condensation energy density}
	
	At arbitrary temperatures, the condensation energy density is given by (hereafter $\hbar=k_B=1$ except for the final expressions)
	\begin{multline}
		\label{Econd-def}
		\frac{\eps_\text{cond}(T)}{N_0}
		=
		- \frac{\Delta^2}{\lambda}
		+
		4\pi T \sum_{\eps>0}^{\omega_D}
		\left( \mathfrak{E} - \eps \right)
		\\{}
		=
		2\pi T \sum_{\eps>0}
		\left(
		2 \mathfrak{E} - 2 \eps - \frac{\Delta^2}{\mathfrak{E}}
		\right) ,
	\end{multline}
	where summation goes over the Matsubara energies $\eps=2\pi T(n+1/2)$, $\mathfrak{E}=\sqrt{\Delta^2+\eps^2}$, $\lambda$ is the dimensionless Cooper-channel interaction constant, $\omega_D$ is the Debye frequency, and $\Delta(T)$ is the BCS value of the order parameter.
	
	In the limit $T\to T_c$, the term $\Delta^2$ vanishes, and keeping $\Delta^4$ we recover the GL expression written below Eq.\ \eqref{FLAMH}.
	In the low-temperature limit, Eq.\ \eqref{Econd-def} reproduces the known result:
	$\eps_\text{cond}(0) = N_0\Delta^2(0)/2$, where
	$\Delta(0)=1.76\, k_BT_c$.
	
	At intermediate $T$ one can use an approximate expression \eqref{eq4}.
	This expression correctly reproduces both the low-$T$ and $T\to T_c$ asymptotics, with the overall error  less than 1\% in the whole temperature range.

	\subsection{Coherence length}
	
	In the GL region, the usual definition of the coherence length for dirty superconductors is $\xi_\text{GL}^2(T) = \pi\hbar D/8k_B(T_c-T)$. It comes from investigating the linear term in the GL equation:
	$(1 + \xi^2 \nabla^2) \Delta + \dots$
	
	At arbitrary temperatures we define $\xi(T)$ in a similar way as the length scale in the equation for small variations of the pairing amplitude, $\delta\Delta$, from its mean-field value, assuming they change slowly in space. They are determined by the fluctuation propagator \cite{FS2016}
	\be
	\label{L-BCS-0}
	L^{-1}(\bq)
	=
	4\pi T
	\sum_{\eps>0}
	\left[
	\frac{1}{2 \mathfrak{E}}
	-
	\frac{\eps^2}{\mathfrak{E}^2(Dq^2 + 2\mathfrak{E})}
	\right] .
	\ee
	At small $q$ one can expand
	\be
	L^{-1}(\bq)
	=
	L^{-1}(0)
	+
	c(\Delta,T) Dq^2
	+ o(q^2) ,
	\ee
	where
	\be
	\label{invL}
	L^{-1}(0)
	=
	2\pi T
	\sum_{\eps>0}
	\frac{\Delta^2}{\mathfrak{E}^3}
	=
	\begin{cases}
		2 (1-T/T_c), & T\to T_c, \\
		1, & T\to 0. \\
	\end{cases}
	\ee
	and
	\be
	\label{c}
	c(\Delta,T)
	=
	\frac{\pi [\Delta+T\sinh(\Delta/T)]}
	{8\Delta T[1+\cosh(\Delta/T)]}
	=
	\begin{cases}
		\pi/8T_c, & T\to T_c, \\
		\pi/8\Delta(0), & T\to 0. \\
	\end{cases}
	\ee
	
	Now for a temperature-dependent coherence length $\xi(T)$ we adopt an operational definition:
	\be
	\label{xi-op}
	\xi^2(T)
	=
	\frac{2\hbar D c(\Delta,T)}{L^{-1}(0)} .
	\ee
	(The factor of 2 in the numerator is related to the fact that our expansion in $\delta\Delta$ is performed near the BCS value rather than near the normal state.) Thus defined $\xi(T)$ coincides with $\xi_\text{GL}(T)$ as $T\to T_c$ and approaches $\sqrt{\pi\hbar D/4\Delta(0)}$ at $T=0$.
	A good approximation of $\xi(T)$ valid for all temperatures (better than 1\% accuracy) is provided by Eq.\ \eqref{eq4}.
	
\begin{ruledtabular}
\begin{table}[t]
  \begin{center}
    \caption{Statistical moments (mean values and corresponding standard deviations) of the distributions of thicknesses $\mu_m$ for different sets of $L$, $M$ and $b$. For the most right column $p$ = 340, $t = u$ = 50; for others $p$ = 20, $t = u$ = 50.}
    \label{tab:tableB1}
       \begin{tabular}{c c c c c}
      & \multicolumn{4}{c}{Set $L$ [$\mu$m]/$M$/$b$ [nm]}\\
      \hline
      \text{Moments} & \text{0.5/50/10} & \text{4/40/100} & \text{20/200/100} & \text{40/40/1000}\\
      \hline
      $\mu_L $[nm] & 5.15 & 5.1 & 5.09 & 5.5\\
      $\sigma\mu_L$ [nm] & 0.49 & 0.51 & 0.5 & 0.34\\
      \end{tabular}
  \end{center}
 \end{table}
\end{ruledtabular}

\section{Thickness uniformity}
	\label{B}
	
\begin{figure}[t]
		\centerline{\includegraphics[width=0.48\textwidth]{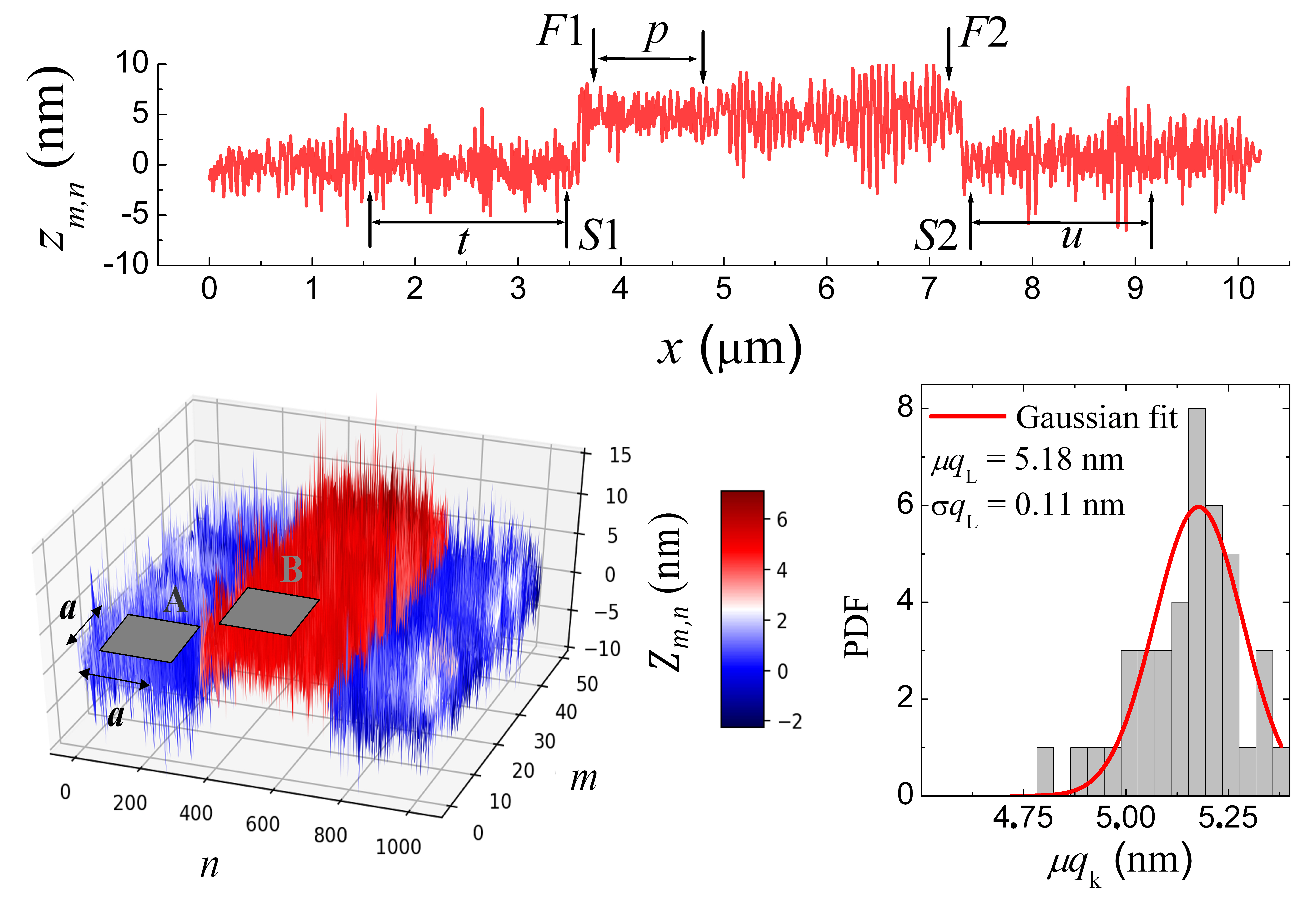}}
		\caption{Top panel: Height profile of a representative levered scan. Arrows mark the borders of areas used for the definition of the strip thickness (Eq.\ \eqref{eqB1}). Labels are the values of the running index $n$ (along the coordinate $x$) corresponding to the borders and the numbers of points involved in the averaging on the strip ($p$) and on the substrate ($t$,\,$u$). Bottom left: AFM image of the portion of the strip with the length $L =$ 500 nm and the squares (not in scale) on the substrate (A) and on the film (B) which were used for alternative definition of the film thickness (Eq.\ \eqref{eqB3}). Bottom right: Histogram (non-normalized PDF) of the statistical distribution of thicknesses $\mu q_k$ and the Gaussian fit (solid line). Fit parameters are listed in the legend.}
		\label{figB1}
	\end{figure}

	For the statistical analysis of the local thickness of our films we prepared a test object in the form of a 4 $\mu$m wide NbN strip with a nominal thickness of 5 nm on the sapphire substrate. The strip had a length of 40 $\mu$m and was surrounded on both sides by 1 mm wide substrate fields. These areas appear after ion etching of the deposited film. To access additional roughness, which may appear as the result of ion etching, we characterized the substrate before film deposition. The strip was prepared by means of the same technological route as the route which was used to prepare nanowires for the main experiment. The strip was scanned with an atomic force microscope (WITec alpha 300 RA) across the strip midline. All scans had the same total length of 10 $\mu$m and were acquired with the resolution $r$ = 10 nm (distance between points in individual scans). Distance between adjacent scans along the strip varied from 10 nm to 1 $\mu$m depending on the length of the studied part of the strip. We levered each scan, i.e. programmatically eliminated the tilt of the specimen with respect to the instrument reference plane. We then evaluated the mean thickness for each scan as
		\begin{equation}
		\mu_m =
		\frac{1}{p}\sum _{n=F1}^{F1+p} [z_{m,n}-\frac{1}{t+u}
		(\sum _{n=S1-t}^{S1} z_{m,n} +
		\sum _{n=S2}^{S2+u} z_{m,n})] .
		\label{eqB1}
	    \end{equation}
	Here $n$ and $m$ are the running indices, which define the number of a point in the scan and the number of the scan, respectively,  $z_{m,n}$ is the levered height of the cantilever, $p<F2-F1$ and $t$ and $u$ are the total numbers of points on the film and on the substrate (on each side of the strip), which were accounted for averaging. Points with numbers $n=F1$ and  $n=F2$ on the strip and  $n=S1$ and  $n=S2$ on the substrate are spaced 100 nm apart from the nearest strip edge in order to avoid edge effects. This configuration is shown in Fig.\ \ref{figB1}.

The mean thickness, $\mu_L$, for the part of the strip with the length $L$  and the standard deviation (STD), $\sigma\mu_L$, in the distribution of mean scan thicknesses $\mu_m$ were computed according standard definitions as
	\begin{equation}
	\mu_L = \frac{1}{M}\sum _{m=1}^{M} \mu_m ,
\qquad
	\sigma\mu_L = \sqrt{\frac{1}{M}
	\sum _{m=1}^{M} (\mu_m-\mu_L)^2} ,
	\label{eqB2}
	\end{equation}
where $M$ is the number of scans within the length $L$ and $L/M$ = $b$  represents the distance between adjacent scans. Values $\mu_L$ and $\sigma\mu_L$ for different combinations of parameters $L$, $M$ and $b$ are listed in Table\ \ref{tab:tableB1}.

Alternatively, to get thickness variations relevant to the hot spot in the main experiment we defined the thickness for a square on the strip adjacent to the strip edge. We chose the square with the side length of 100 nm close to the diameter of the hot spot. The distance between scans was in this case equal to the scan resolution $b = r =$ 10 nm and the strip length was 0.5 $\mu$m. The number of points in the square was $p\,N$ = 100 where $N$ = 10 is the number of involved scans and $p=N$. The reference for the film thickness was defined as the mean height for the square on the substrate with the same size and at the same position along the strip.  The configuration is shown in Fig.\ \ref{figB1}.

 The mean thickness for the square on the strip with the number $k$ was computed as
    \begin{equation}
    \mu q_k = \frac{1}{a^2}\sum _{m=k}^{k+a}
    \sum _{n=F1}^{F1+a} \left(z_{m,n} -
    \frac{1}{a^2}\sum _{m=k}^{k+a} \sum _{n=F1}^{F1+a} z_{m,n} \right) .
	\label{eqB3}
	\end{equation}
For the part of the strip with the length $L$, the mean value of the thickness, $\mu q_L$,  and the standard deviation, $\sigma\mu q_L$, in the distribution of square thicknesses $\mu q_k$ were computed for a set of squares offset by 10 nm in a way presented by Eq.\ \eqref{eqB2} with $k$ running from 1 to $L/r - N$.

The histogram, i.e. non-normalized probability density function (PDF) of the statistical distribution of $\mu q_k$ for $L$ = 0.5 $\mu$m along with the Gaussian fit is shown in Fig.\ \ref{figB1} (bottom right).  Numerically computed moments of this distribution, $\mu q_L$ = 5.15 nm and $\sigma\mu q_L$ = 0.13 nm, are close to values 5.19 nm and 0.11 nm returned by the best Gaussian fit.

Data in Table\ \ref{tab:tableB1} show that $\sigma\mu q_L$ does not systematically depend on the distance between scans, $b$,  and the length, $L$, of the chosen part of the strip. However, STD decreases with the increase of the number of scans, $M$, or the number of points in the scan, $p$, which are involved into averaging. These findings evidence the uniformity of the strip thickness for lengths up to 40 $\mu$m and the absence of any correlations in the strip thickness for distances larger than 10 nm. We have additionally computed autocorrelation function for the thickness following the formalism presented in Ref.\ \cite{Rod 2020} (Eq. A1) and extracted the correlation radius $R_d <$ 10 nm.

Statistical distribution of mean square thicknesses $\mu q_k$ has STD equal to 0.13 nm which is almost three times less than the smallest standard deviation (0.37 nm) in the distribution of mean scan thicknesses $\mu_k$ (Table\ \ref{tab:tableB1}, the most right column). Bearing in mind that the total numbers of averaged points on the strip and on the substrate are approximately the same for both algorithms (scans and squares), we attribute the drastic improvement in the STD for the second square-algorithm to a smaller mean square distance between points on the strip and on the substrate. We therefore conclude that the evaluated STD in the local strip thickness $\sigma\mu q_L$ = 0.13 nm is the upper boundary for the actual variations in the mean strip thickness within squares.

 Standard mean square roughness \cite{34}, $S_q$, for squares on the substrate and on the strip was 2.65$\pm$0.23 nm and 3.08$\pm$0.38 nm, respectively. The $S_q$ value measured on the substrate before film deposition was 2.84 nm, which is close to the roughness of squares. Finally, the instrument noise of 2.0 nm, which was defined as the $S_q$ value for 40000 acquisitions within an area less than $r^2$, is slightly less  but comparable to the $S_q$ values for the squares. Assuming that the instrument noise and the substrate roughness are statistically independent, we obtain the true substrate roughness 1.74 nm. We note that STM surface roughness for our films of $\approx$ 0.1 nm \cite{Noat 2013, Rod 2020} was less than the thickness variations found in the present study that is most likely due to a larger noise in the used AFM instrument.


\begin{thebibliography}{99}
		
		\bibitem{1} L. G. Aslamazov and A. I. Larkin, The influence of fluctuation pairing of electrons on the conductivity of normal metal, Phys. Lett. \textbf{26A} 238, (1968).
		
		\bibitem{2} M. Caloz, B. Korzh, N. Timoney, M. Weiss, S. Gariglio, R. J. Warburton, C. Schonenberger, J. Renema, H. Zbinden, and F. Bussi\`eres, Optically probing the detection mechanism in a molybdenum silicide superconducting nanowire single-photon detector, Appl. Phys. Lett. \textbf{110}, 083106 (2017).
		
		\bibitem{3} A. Verevkin, J. Zhang, R. Sobolewski, A. Lipatov, O. Okunev, G. Chulkova, A. Korneev, K. Smirnov, G. N. Gol'tsman and A. Semenov, Detection efficiency of large-active-area NbN single-photon superconducting detectors in the ultraviolet to near-infrared range, Appl. Phys. Lett. \textbf{80}, 4687 (2002).
		
		\bibitem{4} B. A. Korzh, Q-Y. Zhao, S. Frasca, J. P. Allmaras, T. M. Autry, E. A. Bersin, M. Colangelo, G. M. Crouch, A. E. Dane, T. Gerrits, F. Marsili, G. Moody, E. Ramirez, J. D. Rezac, M. J. Stevens, E. E. Wollman, D. Zhu, P. D. Hale, K. L. Silverman, R. P. Mirin, S. W. Nam, M. D. Shaw, K. K. Berggren, Demonstrating sub-3 ps temporal resolution in a superconducting nanowire single-photon detector, arXiv:1804.06839 (2018).
		
		\bibitem{5} M. Sidorova, A. Semenov, H.-W. H\"ubers, A. Kuzmin, S. Doerner, K. Ilin, M. Siegel, I. Charaev and D. Vodolazov, Timing jitter in photon detection by straight superconducting nanowires: Effect of magnetic field and photon flux, Phys. Rev. B \textbf{98}, 134504 (2018).
		
		\bibitem{6} J. P. Allmaras, A. G. Kozorezov, B. A. Korzh, K. K. Berggren, and M. D. Shaw, Intrinsic Timing Jitter and Latency in Superconducting Nanowire Single-photon Detectors, Phys. Rev. Applied \textbf{11}, 034062 (2019).
		
		\bibitem{7} D. Yu. Vodolazov, Minimal Timing Jitter in Superconducting Nanowire Single-Photon Detectors, Phys. Rev. Applied \textbf{11}, 014016 (2019).
		
		\bibitem{8} U. Fano, Ionization yield of radiation. II. The fluctuations in the number of ions, Phys. Rev. \textbf{72}, 26 (1947).
		
		\bibitem{9} A. G. Kozorezov, C. Lambert, F. Marsili, M. J. Stevens, V. B. Verma, J. P. Allmaras, M. D. Shaw, R. P. Mirin, and S. W. Nam, Fano fluctuations in superconducting-nanowire single-photon detectors, Phys. Rev. B \textbf{96}, 054507 (2017).
		
		\bibitem{10} A. N. Zotova and D. Yu. Vodolazov, Intrinsic detection efficiency of superconducting nanowire single photon detector in the modified hot spot model, Supercond. Sci. Technol. \textbf{27}, 125001 (2014).
		
		\bibitem{Cheng 2017} Yuhao Cheng, Chao Gu, and Xiaolong Hu, Inhomogeneity-induced timing jitter of superconducting nanowire single-photon detectors, Appl. Phys. Lett. \textbf{111}, 062604 (2017).
		
		\bibitem{11} A. Semenov, I. Charaev, R. Lusche, K. Ilin, M. Siegel, H.-W. H\"ubers, N. Bralovic, K. Dopf, and D. Yu. Vodolazov, Asymmetry in the effect of magnetic field on photon detection and dark counts in bended nanostrips, Phys. Rev. B \textbf{92}, 174518 (2015).
		
		\bibitem{12} I. Charaev, T. Silbernagel, B. Bachowsky, A. Kuzmin, S. Doerner, K. Ilin, A. Semenov, D. Roditchev, D. Yu. Vodolazov, and M. Siegel, Proximity effect model of ultranarrow NbN strips, Phys. Rev. B \textbf{96}, 184517 (2017).
		
		\bibitem{13} J. S. Langer and V. Ambegaokar, Intrinsic Resistive Transition in Narrow Superconducting Channels, Phys. Rev. \textbf{164}, 498 (1967).
		
		\bibitem{14} D. E. McCumber and B. I. Halperin, Time Scale of Intrinsic Resistive Fluctuations in Thin Superconducting Wires, Phys. Rev. B \textbf{1}, 1054 (1970).
		
		\bibitem{15} J. Bardeen, Critical fields and currents in superconductor, Rev. Mod. Phys. \textbf{34}, 667 (1962).
		
		\bibitem{16} Fundamentals of Photonics, Bahaa E. A. Saleh, Malvin Carl Teich, John Wiley $\&$ Sons, Inc, USA 1991, ISBN: 0-471-83965-5.
		
		\bibitem{Kermann 2007} A. J. Kerman, E. A. Dauler, J. K. W. Yang, K. M. Rosfjord, V. Anant, K. K. Berggren, G. N. Gol’tsman, and B. M. Voronov, Appl. Phys. Lett. \textbf{90} (10), 101110 (2007).
		
		\bibitem{Rod 2020} C. Carbillet, V. Cherkez, M. A. Skvortsov, M. V. Feigel’man, F. Debontridder, L. B. Ioffe, V. S. Stolyarov, K. Ilin, M. Siegel, C. Noûs, D. Roditchev, T. Cren, and C. Brun, Spectroscopic evidence for strong correlations between local superconducting gap and local Altshuler-Aronov density of states suppression in ultrathin NbN films, Phys. Rev. B  \textbf{102}, 024504 (2020).

		\bibitem{Landau} Equilibrium and non-equilibrium statistical mechanics, Radu Balescu, John Willey \& Sons, New York, 1975 (pp. 131-135); Statistical Physics, Part 1, L.D. Landau and E.M. Lifshitz  (3rd ed.), Pergamon Press (1985).
		
		\bibitem{U_FLUCT} K. M. van Vliet and H. Menta,Theory of Transport Noise in Semiconductors, Phys. Stat. Sol. \textbf{106}106, 11 (1981).
		
		\bibitem{TES} K.D. Irwin and G.C. Hilton, Transition edge sensors, Topic of Applied Physics \textbf{99}, 63-149 (2005).
		
		\bibitem{HEB} B.S. Karasik and A.I. Elantiev, Appl. Phys. Lett. \textbf{68} (6), 853 (1996).
		
		\bibitem{MBL} P.L. Richards, J. Appl. Phys. \textbf{76} (1), 1 (1994).
		
		\bibitem{19} A. V. Semenov, P. A. Krutitsky, and I. A. Deviatov, Microscopic theory of phase slip in a narrow dirty superconducting strip, Pis'ma v ZhEFT \textbf{92}, 842 (2010) [JETP Letters \textbf{92}, 762 (2010)].
		
		\bibitem{Marychev2017} P. M. Marychev and D. Yu. Vodolazov, Threshold fluctuations in a superconducting current-carrying bridge Supercond. Sci. Technol. \textbf{30}, 075008 (2017).

		\bibitem{Vodolazov2012} D. Yu. Vodolazov, Saddle point states in two-dimensional superconducting films biased near the depairing current, Phys. Rev. B \textbf{85}, 174507 (2012).	
		
		\bibitem{20} L. N. Bulaevskii, M. J. Graf, C. D. Batista and V. G. Kogan, Vortex-induced dissipation in narrow current-biased thin-film superconducting strips, Phys. Rev. B \textbf{83}, 144526 (2011).
		
		\bibitem{21} A. Semenov, B. G\"unther, U. B\"uttger, H.-W. H\"ubers, H. Bartolf, A. Engel, A. Schilling, K. Ilin, M. Siegel, R. Schneider, D. Gerthsen, and N. A. Gippius, Optical and transport properties of ultrathin NbN films and nanostructures, Phys. Rev. B \textbf{80}, 054510 (2009).
		
		\bibitem{22} D. T. Gillespie, The mathematics of Brownian motion and Johnson noise, Am. J. Phys. \textbf{64}, 225 (1995).
		
		\bibitem{23} A. Lipton and V. Kaushansky, On the First Hitting Time Density of an Ornstein-Uhlenbeck Process, arXiv:1810.02390v2 [q-fin.CP] 10 Oct 2018.
		
		\bibitem{24} L. Sacerdote, Asymptotic behaviour of Ornstein-Uhlenbeck first-passage-time density through periodic boundaries, Applied Stochastic Models and Data Analysis \textbf{6}, 53 (1990).
		
		\bibitem{25} R. J. Martin, M. J. Kearney and R. V. Craster, Long- and short-time asymptotics of the first-passage time of the Ornstein-Uhlenbeck and other mean-reverting processes, arXiv:1810.13010v1 [math-ph] 30 Oct 2018.
		
		\bibitem{Sidor 2020} Mariia Sidorova, Alexej Semenov, Heinz-Wilhelm Hübers, Konstantin Ilin, Michael Siegel, Ilya Charaev, Maria Moshkova, Natalia Kaurova, Gregory N. Goltsman, Xiaofu Zhang, and Andreas Schilling, Electron energy relaxation in disordered superconducting NbN films, Phys. Rev. B \textbf{102}, 054501 (2020).

		\bibitem{LosAlam} Shi-Zeng Lin, Oscar Ayala-Valenzuela, Ross D. McDonald, Lev N. Bulaevskii, Terry G. Holesinger, Filip Ronning, Nina R. Weisse-Bernstein, Todd L. Williamson, Alexander H. Mueller, Mark A. Hoffbauer, Michael W. Rabin, and Matthias J. Graf, Characterization of the thin-film NbN superconductor for single-photon detection by transport measurements Phys. Rev. B  \textbf{87}, 184507 (2013).
		
		\bibitem{32} K. Yakushiji, S. Mitani, K. Takanashi, J.-G. Ha, and H. Fujimori, Composition dependence of particle size distribution and giant magnetoresistance in Co-Al-O granular films, J. Magn. Magn. Mat. \textbf{212}, 75 (2000).		
		
		\bibitem{27} L. Maingault, M. Tarkhov, I. Florya, A. Semenov, R. Espiau de Lama\"estre, P. Cavalier, G. Gol'tsman, J.-P. Poizat, and J.-C. Vill\'egier, Spectral dependency of superconducting single photon detectors, J. Appl. Phys. \textbf{107}, 116103 (2010).
		
		\bibitem{28} S. Ferrary, V. Kovalyuk, W. Hartmann, A. Vetter, O. Kahl, C. Lee,  A. Korneev,  C. Rockstuhl, G. Goltsman and W. Pernice, Hot-spot relaxation time current dependence in niobium nitride waveguide-integrated superconducting nanowire single-photon detectors, Optics Express \textbf{25}, 8739 (2017).
		
		\bibitem{29} D. L. Goodstein, A. W. Harter, and T. C. P. Chui, Heat capacity of a current carrying superconductor, Phys. Lett. A \textbf{245}, 477 (1998).
		
		\bibitem{FS2016}
		Ya. V. Fominov and M. A. Skvortsov,
		Subgap states in disordered superconductors with strong magnetic impurities,
		Phys. Rev. B \textbf{93}, 144511 (2016).
		
		\bibitem{34}
		Surfaces and their Measurement, David J. Whitehouse, Ed. Taylor Hobson Ltd, London, 2002 (p. 84, Eq. 3.20).
		
		\bibitem{Noat 2013} Y. Noat, V. Cherkez, C. Brun, T. Cren, C. Carbillet, F. Debontridder, K. Ilin, M. Siegel, A. Semenov, H.-W. Hübers, and D. Roditchev, Phys. Rev. B \textbf{88}, 014503 (2013).
		
	\end{thebibliography}
\end{document}